\documentclass[a4paper,fleqn,usenatbib]{mnras}

\usepackage{newtxtext,newtxmath}

\usepackage[T1]{fontenc}
\usepackage{ae,aecompl}

\usepackage{graphicx}	
\usepackage{amsmath}	
\usepackage{amssymb}	

\title[Dependence of stellar properties on density]{The dependence of stellar properties on initial cloud density}

\author[M. O. Jones et al.]{
Michael O. Jones$^{1}$\thanks{E-mail: mjones@astro.ex.ac.uk (MOJ); mbate@astro.ex.ac.uk (MRB)} and Matthew. R. Bate$^{1}$
\\
$^{1}$School of Physics and Astronomy, University of Exeter, Stocker Road, Exeter EX4 4QL\\
}

\date{Accepted by MNRAS}

\pubyear{2017}

\begin{document}
\label{firstpage}
\pagerange{\pageref{firstpage}--\pageref{lastpage}}
\maketitle

\begin{abstract}

We investigate the dependence of stellar properties on the initial mean density of the molecular cloud in which stellar clusters form using radiation hydrodynamical simulations that resolve the opacity limit for fragmentation. We have simulated the formation of three star clusters from the gravitational collapse of molecular clouds whose densities vary by a factor of a hundred. As with previous calculations including radiative feedback, we find that the dependence of the characteristic stellar mass, $M_{\rm c}$, on the initial mean density of the cloud, $\rho$, is weaker than the dependence of the thermal Jeans mass. However, unlike previous calculations, which found no statistically significant variation in the median mass with density, we find a weak dependence approximately of the form $M_{\rm c} \propto \rho^{-1/5}$. The distributions of properties of multiple systems do not vary significantly between the calculations. We compare our results to the result of observational surveys of star-forming regions, and suggest that the similarities between the properties of our lowest density calculation and the nearby Taurus-Auriga region indicate that the apparent excess of solar-type stars observed may be due to the region's low density.

\end{abstract}




\section{Introduction} \label{intro}
Any complete theory of star formation must account not only for the properties and dynamics of individual stellar systems, but also for the distributions of stellar properties and the behaviour of these systems as part of larger ensembles. Although the stellar initial mass function (IMF) has been the primary focus of many studies, a host of other statistical properties must also be explained, such as the star-formation rate and efficiency, the relative numbers of binary and multiple systems as well as the structure and kinematics of the clusters in which star formation takes place. 

A natural approach to this problem is to use numerical models. By simulating an entire star-forming cloud in a single calculation, it is possible to generate a stellar population large enough to examine its statistical properties and compare them with results from observational surveys. Initial attempts at this focussed on the formation of multiple systems \citep{1992Natur.359..207C}, and the distribution of masses in star-forming clusters (\citealt{1998ApJ...501L.205K} \citealt{2000ApJS..128..287K}). \citet{1997MNRAS.285..201B} found that the distribution of masses was primarily determined by variations in stellar accretion rates set by dynamic interactions between protostellar systems, underlining the need to model ensembles of stars in order to correctly predict their properties. 

Most of these early calculations (and many recent calculations) used sink particles with large accretion radii ($\sim 100~\textup{AU}$) to model stars, inhbiting the formation of low-mass stars and brown dwarfs and obscuring the properties of most binaries and discs. Resolving the opacity limit for fragmentation (\citealt{1976MNRAS.176..367L}, \citealt{1976MNRAS.176..483R}) allows the properties of individual stars, binary and multiple systems to be investigated.  The first simulations to do this were those of \citet{2002MNRAS.332L..65B,2002MNRAS.336..705B,2003MNRAS.339..577B}. Using a combination of hydrodynamics and gravitational dynamics with a barotropic equation of state, the simulated clusters were found to over-produce brown dwarfs, with more brown dwarfs being produced than stars. Subsequent calculations using the same method also found that the characteristic (median) stellar mass of the clusters was linearly proportional to the thermal Jeans mass of the initial molecular cloud \citep{2005MNRAS.356.1201B,2005MNRAS.363..363B,2005A&amp;A...435..611J,2006MNRAS.368.1296B}.

It was later shown by several authors that the inclusion of radiative heating prevented the overproduction of brown dwarfs, as the additional thermal support inhibited the collapse of fragments that would otherwise lead to the production of low-mass objects (\citealt{2009MNRAS.392.1363B}; \citealt{2009ApJ...703..131O}). \citet{2012MNRAS.419.3115B} confirmed this with the largest calculation to date to include radiative heating whilst resolving the opacity limit. Using a combination of hydrodynamics, radiative transfer and gravitational dynamics, the calculation produced a stellar population of 183 stars and brown dwarfs, whose statistical properties were in good agreement with observations. 

In addition to demonstrating the effects of radiative heating on the production of brown dwarfs, \citet{2009MNRAS.392.1363B} suggested that it may also weaken the dependence of the median stellar mass on the thermal Jeans mass of the initial molecular cloud. This would provide a possible explanation for the apparent `universality' of the observed IMF \citep{2010ARA&amp;A..48..339B}. Results from two calculations that varied in initial density by a factor of 9 showed no statistically significant difference in the median stellar mass. However, as each calculation only produced populations of $\sim 20$ objects, the possibility of a dependence of the median mass on density that was too weak to be detected with such small numbers of stars remains.

In this paper, we present the results of radiative hydrodynamical calculations of the collapse of turbulent, molecular clouds to form star clusters. The calculations are identical, except for their initial densities, which we vary by up to a factor of 100. We use initial conditions similar to those of \citet{2009MNRAS.392.1363B}, but simulate clouds that are ten times more massive than the original calculations, and equivalent in mass to the those of \citet{2012MNRAS.419.3115B}. The increased scale of these new calculations allows us to test the predictions of \citet{2009MNRAS.392.1363B} using a more thorough statistical analysis than was previously possible.

This paper is structured as follows. Section \ref{method} describes our method, including a brief description of our radiative hydrodynamics method and the initial conditions for the calculations. In section \ref{results}, we present the results of the three cluster calculations, including an analysis of the cluster density and temperature structures, mass functions and multiple system properties. In section \ref{disc}, we discuss the results in the context of previous calculations, and compare them to recent observations of stellar populations. Our conclusions are then presented in section \ref{conc}.

\section{Computational method} \label{method}

The calculations presented in this paper were completed using \texttt{sphNG}, a modified version (\citealt{1995MNRAS.277..362B}, \citealt{2005MNRAS.364.1367W}, \citealt{2006MNRAS.367...32W}) of a three-dimensional smoothed particle hydrodynamics (SPH) code originally developed by \citet{1990ApJ...348..647B,1990nmns.work..269B}, parallelised using both \texttt{OpenMP} and \texttt{MPI}. The code uses a binary tree to calculate the gravitational forces between particles and their nearest neighbours. The smoothing lengths of particles are allowed to vary in time and space, and are set iteratively such that the smoothing length of each particle $h = 1.2(m/\rho)^{1/3}$ where $m$ is the particle mass, and $\rho$ is the particle density (see \citealt{2007MNRAS.374.1347P}). A second-order Runge-Kutta-Fehlberg method \citep{1969NASA...R315} is used to integrate the SPH equations, with individual time-steps for each particle \citep{1995MNRAS.277..362B}. The artificial viscosity prescription given by \citet{1997JCoPh.136...41M} is used, with $\alpha_v$ varying between 0.1 and 1 and $\beta_v = 2\alpha_v$.

\subsection{Radiative transfer \& equation of state} \label{method_RHD}

The calculations are performed using radiation hydrodynamics (\citealt{1984oup..book.....M}, \citealt{2001ApJS..135...95T}), combined with an ideal gas equation of state $p = \rho T_g \mathcal{R} / \mu$, where $T_g$ is the gas temperature, $\mathcal{R}$ is the gas constant and $\mu$ is the mean molecular weight of the gas, set to $\mu = 2.38$. Translational, rotational and vibrational degrees of freedom of molecular hydrogen are accounted for in the thermal evolution of the gas, as well as molecular dissociation of hydrogen and ionisation of both hydrogen and helium, with hydrogen and helium mass fractions of $X = 0.70$ and $Y = 0.28$. We ignore the contribution of metals to the equation of state. 

We implement two-temperature (gas and radiation) radiative transfer using the flux-limited diffusion approximation, as described by \citet{2005MNRAS.364.1367W} and \citet{2006MNRAS.367...32W}, to model the transport of radiation. Energy is generated by doing work on the gas or radiation field, and transferred between the two according to their relative temperatures, as well as the gas density and opacity. We assume that the gas and dust are well-coupled, and that their respective temperatures are the same. A grey opacity is used, set to maximum value of the interstellar grain opacity according to the tables of \citet{1985Icar...64..471P} for low temperatures, and the gas opacity according to the tables of \citet{1975ApJS...29..363A} at high temperatures. 

The molecular clouds have free boundaries. However, to provide a boundary for the radiation field, all SPH particles with a density less than $10^{-21}\textup{~g~cm}^{-3}$ have both their gas and radiation temperatures set to 10 K.

\begin{figure*}
	\includegraphics{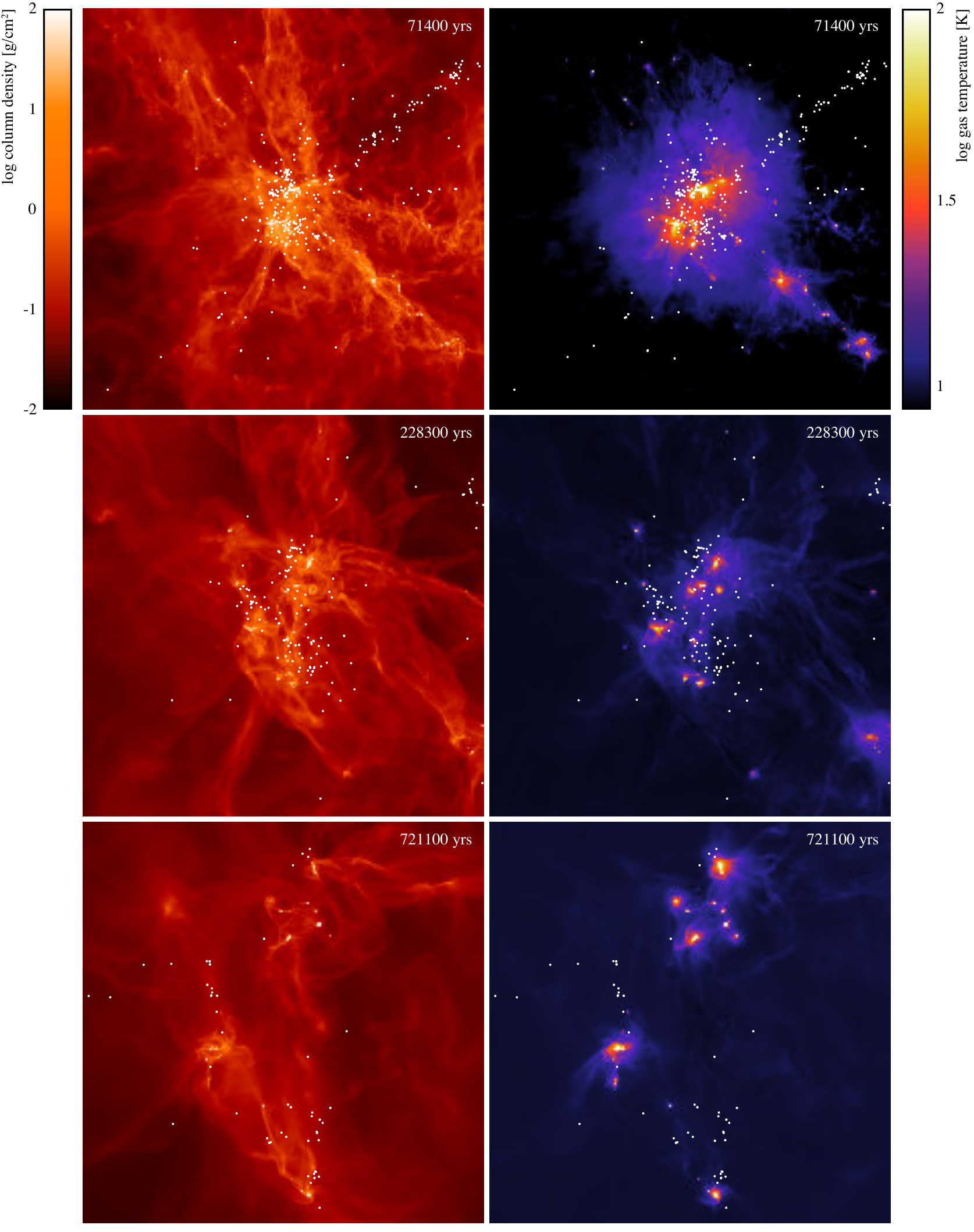}
	\caption{The gas column density (left column) and mass-weight temperature (right column) in the each of the calculations at time $t = 1.2~t_{\textup{ff}}$. The top, middle and bottom rows show the clouds with initial radii of 0.188 pc, 0.404 pc and 0.870 pc, respectively. The dissipation of energy via shocks allows local regions of the clouds to undergo gravitational collapse, producing dense, star-forming cores. Increasing the mean density of the cloud has the effect of producing a more centrally concentrated distribution of cores in the clusters. Each panel is 0.4 pc (82333 AU) across, with the time in years is displayed in the top-right corner. White dots indicate the locations of stars and brown dwarfs.}
	\label{denstemp}
\end{figure*}

\subsection{Sink particles} \label{method_sink}

As discussed in \citet{2012MNRAS.419.3115B}, using radiation hydrodynamics with a realistic equation of state allows us to capture each phase of protostar formation \citep{1969MNRAS.145..271L}. However, as the first hydrostatic core collapses, following the dissociation of molecular hydrogen, the time-step as defined by the Courant-Friedrichs-Levy condition becomes very small, making the required calculation time unfeasibly long, particularly for simulations of the scale presented in this paper.

To alleviate this issue, we use sink particles as described in \citet{1995MNRAS.277..362B}. Once a particle reaches a density of $\rho_{\textup{crit}} = 10^{-5}\textup{~g~cm}^{-3}$, it is replaced with a sink particle with an accretion radius, $R_{\text{acc}}=0.5$ AU. The sink particles accrete SPH particles that pass within the accretion radius and the mass of any accreted particles is subsequently added to the sink mass. Sink particles are permitted to merge if they pass within 0.015 AU of each other, as young protostars are expected to be larger than the Sun (e.g. \citealt{2009ApJ...691..823H}). They provide no hydrodynamic pressure, and do not emit radiation.

By omitting radiation from inside sink particles, the protostellar luminosity is underestimated by a factor of $\approx R_{\text{acc}}/R_*$. \citet{2012MNRAS.419.3115B} gave a detailed discussion of the potential effects of this underestimated luminosity. Comparing calculations using different sink accretion radii, \citeauthor{2012MNRAS.419.3115B} concluded that using $R_{\text{acc}}=0.5$ AU provides sufficient protostellar luminosity to suppress the majority of the anomalous fragmentation that occurs without radiative transfer and that using smaller accretion radii did not reduce the amount of fragmentation much further. Using this approach, Bate was able to obtain a star to brown-dwarf ratio that was in good agreement with observed values. We take the same approach, and discuss the implications for our results in section \ref{disc}.

\subsection{Initial conditions} \label{method_initial}

We simulate three turbulent, uniform density gas clouds, using the same initial conditions as \citet{2012MNRAS.419.3115B}, but varying the initial densities by factors of 10, 1, and 0.1 (i.e. over a range of a factor of 100). This is achieved by constructing uniform spheres containing $500~\textup{M}_{\odot}$ of gas, with radii of 0.188 pc, 0.404 pc and 0.870 pc. The initial mean densities of the sphere are therefore $1.2 \times 10^{-19}\textup{~g~cm}^{-3}$, $1.2 \times 10^{-20}\textup{~g~cm}^{-3}$ and $1.2 \times 10^{-21}\textup{~g~cm}^{-3}$, corresponding to number densities of $\sim 300 \textup{~cm}^{-3}$, $\sim 3 \times 10^3 \textup{~cm}^{-3}$ and $\sim 3 \times 10^4 \textup{~cm}^{-3}$, respectively. The initial temperatures of the spheres are set at 10.3 K. While real molecular clouds of different densities may be expected to have somewhat different temperatures (e.g. due to different levels of extinction), in this study we wish to vary only one parameter at a time so we do not vary the initial temperature.

An initial `supersonic', turbulent velocity field was applied to each cloud, in the manner described by \citet{2001ApJ...546..980O} and \citet{2003MNRAS.339..577B}. To do this, a divergence-free, random Gaussian field was generated with a power spectrum of $P \propto k^{-4}$, where $k$ is the wavenumber, giving a velocity dispersion which varies with distance $\lambda$ as $\sigma(\lambda) \propto \lambda^{-1/2}$ in three dimensions, in agreement with the Larson scaling relations for molecular clouds \citep{1981MNRAS.194..809L}. The field was generated on a $128^3$ grid and the velocities interpolated from the grid. The field was normalised such that the total kinetic energy was equal to the gravitational potential energy of the sphere. The initial free-fall times were $1.90 \times 10^{12}~\textup{s}$ ($6.0\times 10^4$ yr), $6.00 \times 10^{12}~\textup{s}$ ($1.90\times 10^5$ yr) and $1.90 \times 10^{13}~\textup{s}$ ($6.0 \times 10^5$ yr). 

The random seed used to realise the velocity field used in these calculations differed from that used in the calculation by \citet{2012MNRAS.419.3115B}. If the statistical properties of stars and brown dwarfs depend only on global quantities, the results should be unaffected by a change in the random seed. Since we reproduce the same initial conditions as the 0.404 pc case in \citet{2012MNRAS.419.3115B}, by using different random seeds we are able to test this hypothesis. 

\subsection{Resolution} \label{method_res}
To correctly model fragmentation down to the opacity limit, the local Jeans mass must be resolved (\citealt{1997MNRAS.288.1060B}; \citealt{1997ApJ...489L.179T}; \citealt{1998MNRAS.296..442W}; \citealt{2000ApJ...528..325B}; \citealt{2006A&amp;A...450..881H}). As in \citet{2012MNRAS.419.3115B,2014MNRAS.442..285B}, we use $3.5 \times 10^{7}$ particles to model the $500~\textup{M}_{\odot}$ clouds, giving a mass resolution of $1.4 \times 10^{-5}~\textup{M}_{\odot}$ per particle ($7 \times 10^{4}$ particles per $~\textup{M}_{\odot}$).

\begin{figure}
	\includegraphics{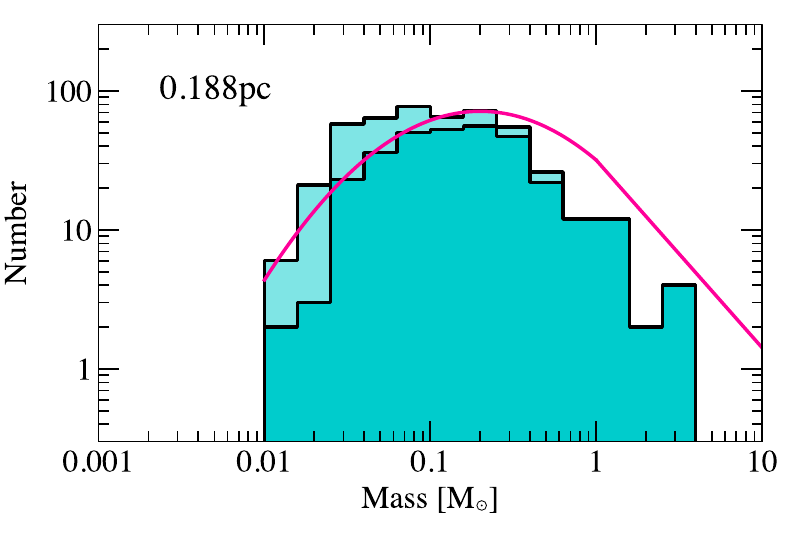} 
	\includegraphics{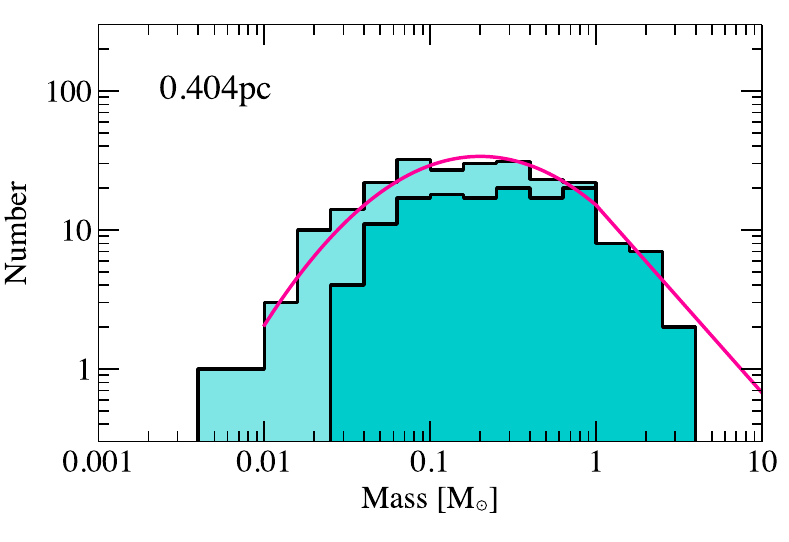} 
	\includegraphics{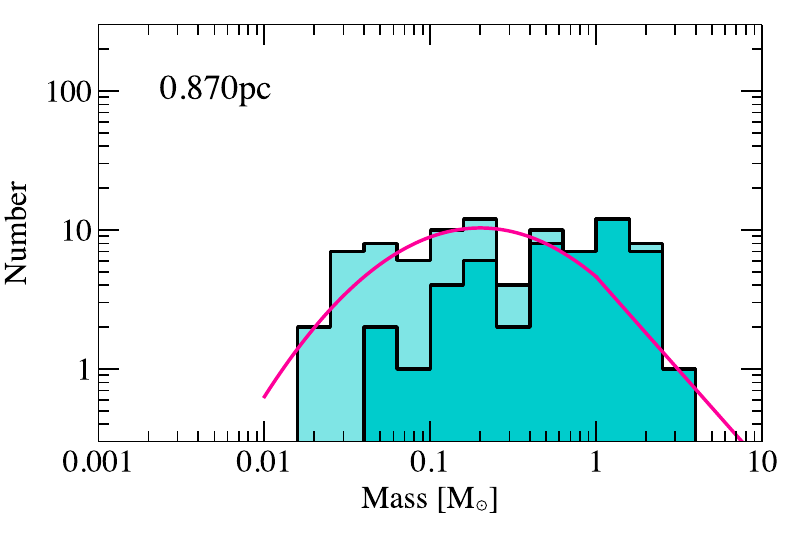} \vspace{-0.2cm}
	\caption{\label{IMF_hist} The distribution of stellar and brown dwarf masses in the each of the calculations at time $t = 1.2~t_{\textup{ff}}$. The dark shaded areas represent objects that have are still accreting when the calculations were stopped (defined as accreting at a rate if less than $10^{-7} ~\textup{M}_{\odot} ~\textup{yr}^{-1}$, and the light shaded areas show those that have stopped accreting. We also plot the parameterisation of the observed Galactic IMF by \citet{2005ASSL..327...41C} in magenta for comparison. There is a weak, but clear dependence of the shape of the IMFs on the mean cloud density. Increasing the mean density of the cloud appears to increase the relative number of low-mass stars produced, and decrease the number of intermediate-mass stars. Decreasing the cloud mean density has the inverse effect.}
\end{figure}

\section{Results} \label{results}
We present results from the three radiation hydrodynamical simulations of star-cluster formation, varying in initial density by up to a factor of 100. The calculations were run for 1.20 initial cloud free-fall times, by which time the 0.188 pc, 0.404 pc and 0.870 pc clouds had produced 474, 233 and 87 stars and brown dwarfs, with total masses of $91.2~\textup{M}_{\odot}$, $43.2~\textup{M}_{\odot}$ and $13.2~\textup{M}_{\odot}$, respectively. 

We follow a similar analysis to that of \citet{2012MNRAS.419.3115B} and \citet{2014MNRAS.442..285B}. However, in the interest of brevity, we focus our discussion on the final state of each cluster and their statistical properties. In section \ref{results_denstemp}, we describe the  density and temperature structure of each cluster. In section \ref{results_IMF}, we examine the stellar mass distributions of the clusters, and compare them to analytical models and previous results. Sections \ref{results_mult}, \ref{results_sep} and \ref{results_MR} give details of the multiple systems produced each calculation, comparing the multiplicities, separation distributions and mass distributions. We omit more detailed aspects of the multiple system properties, such as orbit orientations and eccentricities, as well as the time evolution and kinematic properties of the clusters.

\subsection{Density \& temperature distributions} \label{results_denstemp}

Fig. \ref{denstemp} shows the gas column-density and mass-weighted temperature distributions at the end of each of the calculations. In all cases the initial `turbulent' velocity field results in filamentary structure. Dense cores form in regions where filaments meet, as shocks dissipate the turbulent energy initially supporting the cloud, allowing local regions of the cloud to collapse (see \citealt{2003MNRAS.343..413B}, \citealt{2014ApJ...791..124G}). These cores become the primary sites of star-formation in the cloud, heating the surrounding regions as gas falls into the potential well of the core and is accreted by the protostars within. 

The higher the initial density of the molecular cloud, the greater the initial number of Jeans masses contained within it. This leaves the gas more unstable to fragmentation. The effect is clear in the high-density 0.188 pc calculation, as large regions undergo rapid fragmentation, resulting in a single star-forming core that encompasses much of the cluster. The lower density of the 0.404 pc and 0.870 pc calculations gives rise to more dispersed regions of star-formation, as the initial Jeans length is larger and the distance between regions unstable to collapse is increased. 

The level of instability to collapse is also reflected in the star-formation rate of the clusters, which increases with higher initial cloud density. The mean star-formation rates over the duration of the calculations are $1.26 \times 10^{-3}~\textup{M}_{\odot}~\textup{yr}^{-1}$ for the 0.188 pc cloud; $0.19 \times 10^{-3}~\textup{M}_{\odot}~\textup{yr}^{-1}$ for the 0.404 pc cloud; and $0.02 \times 10^{-3}~\textup{M}_{\odot}~\textup{yr}^{-1}$ for the 0.870 pc cloud.

As the temperature of a region heated by an embedded source increases with luminosity and decreases with distance from the source, regions where star-formation is more heavily concentrated are heated to higher temperatures. The higher gas density in these regions also increases the optical depth, trapping radiation emitted by the accreting protostars and heating the star-forming core. The high star formation rate and short distances between protostars in the 0.188 pc calculation result in a large heated region approximately 0.2 pc in diameter with temperatures of $\approx 20-50~\textup{K}$. The 0.404 pc and 0.870 pc calculations also contain heated regions. However, at these lower densities, the cores become more isolated and the heated regions are smaller.

\begin{figure}
	\includegraphics{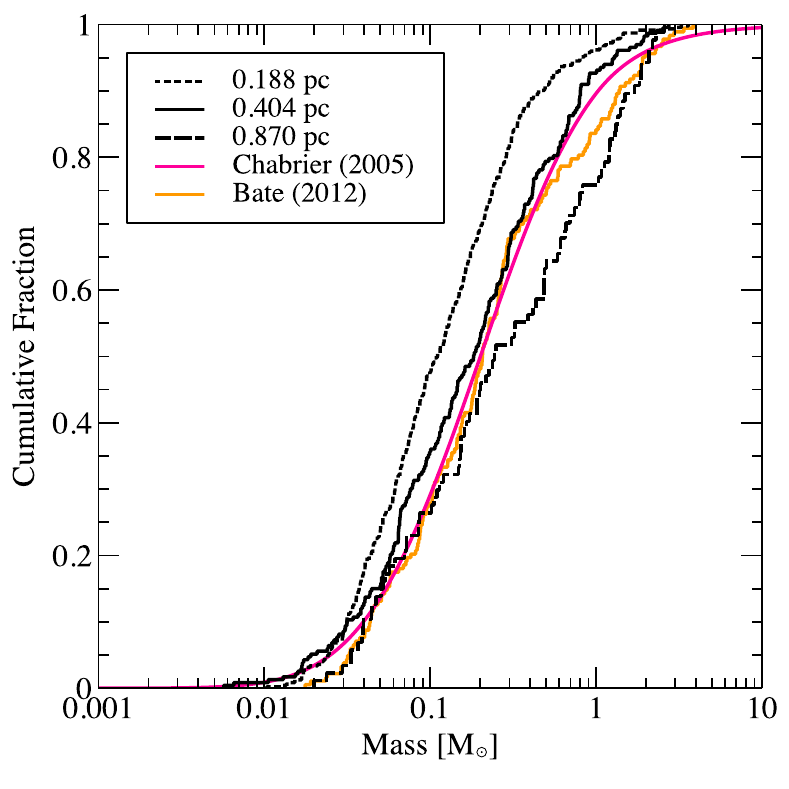}
	\caption{The cumulative distribution of stellar and brown dwarf masses in the each of the calculations at time $t = 1.2~t_{\textup{ff}}$. Results from the 0.188 pc calculation are shown by the dotted line, the 0.404 pc results by the solid line, and the 0.870 pc results by the dashed line. We also plot the \citet{2005ASSL..327...41C} parameterisation of the IMF in magenta and the results of \citet{2012MNRAS.419.3115B} in orange for comparison. Here the differences between the mass distributions produced by the calculations are clear. The results of the 0.404 pc calculation are in good agreement with both the \citet{2005ASSL..327...41C} and \citet{2012MNRAS.419.3115B} mass functions, however there is an excess of low-mass objects formed in the 0.188 pc calculation, and an excess of intermediate-mass stars in the 0.870 pc calculation.}
	\label{IMF_cum}
\end{figure}

\subsection{The initial mass function} \label{results_IMF}
Fig. \ref{IMF_hist} shows the differential mass functions at the end of each of the three calculations. We compare each with the parameterisation of the observed Galactic IMF by \citet{2005ASSL..327...41C}. 

It is clear that increasing the initial density of the cloud results in the production of more low-mass stars and fewer intermediate-mass stars, whereas the converse is true for decreasing density. This results in a shift in the characteristic (median) stellar mass, which is $0.098~\textup{M}_{\odot}$ for the 0.188 pc calculation, $0.182~\textup{M}_{\odot}$ for the 0.404 pc calculation, and $0.238~\textup{M}_{\odot}$ for the 0.870 pc calculation. If a relationship of the form $M_{\rm c} \propto \rho^{-n}$ is assumed, then these shifts imply exponents of 0.27 between the  highest density and intermediate density calculations, and 0.12 between the lowest density and intermediate density calculations. The mean exponent over all three calculations is 0.19. 

The mass function also becomes noticeably `broader' as the initial density of the cloud decreases. This is reflected in an increase in the standard deviations of the distributions, which are 0.469 dex for the 0.188 pc calculation, 0.543 dex for the 0.404 pc calculation, and 0.594 dex for the 0.870 pc calculation. 

As such, although the results of the 0.404 pc calculation are in good agreement with the parameterisation by \citet{2005ASSL..327...41C}, the IMFs produced by increasing or decreasing the initial cloud density from the 0.404 pc case differ significantly, with the 0.188 pc calculation producing too many low-mass stars and brown dwarfs, and the 0.870 pc calculation producing too many intermediate-mass stars.

\begin{figure}
	\includegraphics{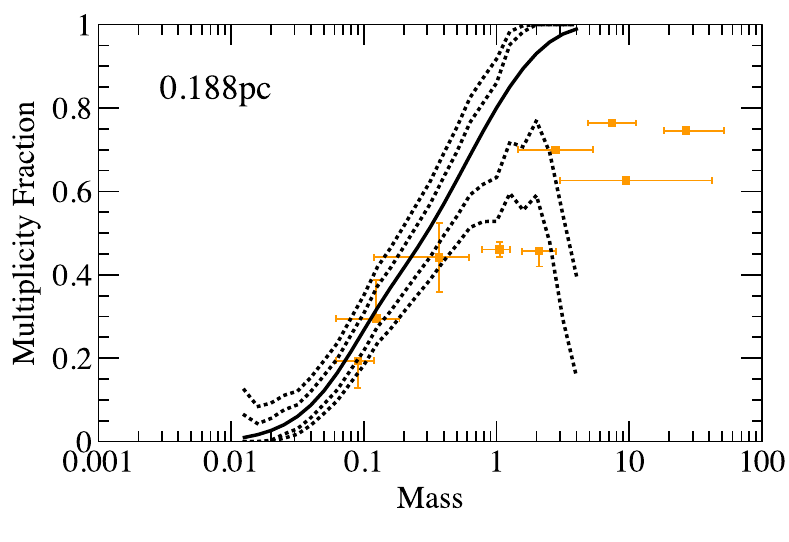}
	\includegraphics{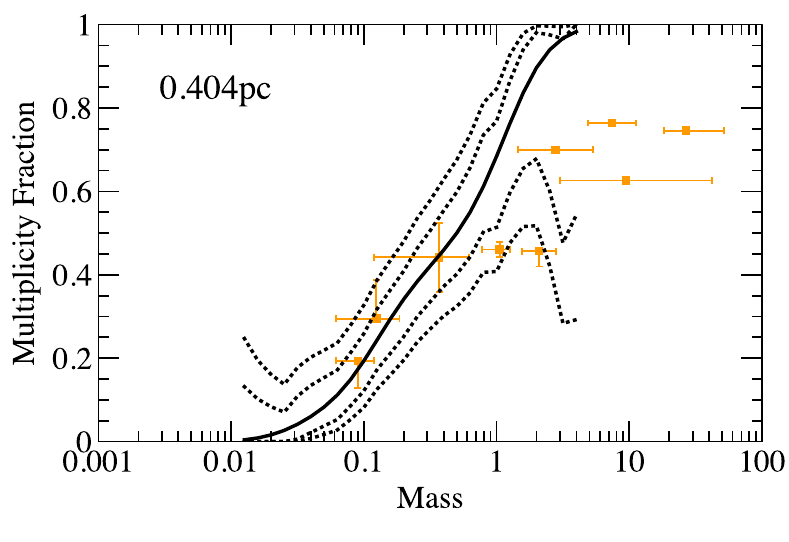}
	\includegraphics{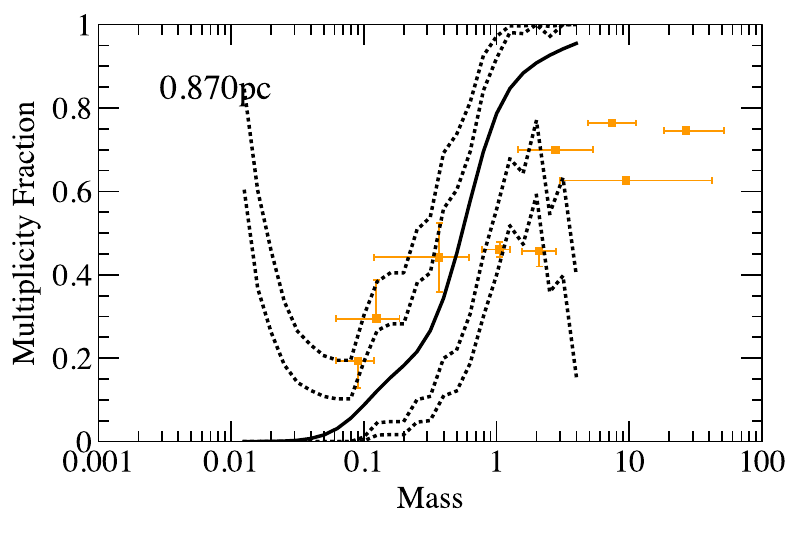}
	\caption{The multiplicity fraction as a function of primary mass in the each of the calculations at time $t = 1.2~t_{\textup{ff}}$. The solid line shows the multiplicity fraction, calculated using a moving lognormal average, with the dotted lines representing the $1 \sigma$ and $2 \sigma$ confidence intervals around the solid line. Filled squares with error bars show the observed multiplicity fractions from surveys by \citet{2003ApJ...587..407C}, \citet{2006AJ....132..663B}, \citet{1992ApJ...396..178F}, \citet{2010ApJS..190....1R}, \citet{1991A&amp;A...248..485D}, \citet{2007A&amp;A...474...77K}, \citet{2013MNRAS.436.1694R},\citet{1999NewA....4..531P} and \citet{1998AJ....115..821M}. Each of the calculations produces and increasing multiplicity with primary mass, in qualitative agreement with observations. There is, however, an increased multiplicity for solar-type stars across all of the calculations when compared to observations by \citet{2010ApJS..190....1R}. This may be due to the young age of the stellar population }
	\label{multfrac_all}
\end{figure}

Fig. \ref{IMF_cum} shows the cumulative mass function for each of the calculations. Again, we plot the parameterisation by \citet{2005ASSL..327...41C} and also include the result from \citet{2012MNRAS.419.3115B}  for comparison. The overproduction of low-mass objects in the 0.188 pc calculation is even more apparent here, with stars of mass $\leq 0.1 ~\textup{M}_{\odot}$ making up $48\%$ of the produced population, compared with $36\%$ in the 0.404 pc calculation. Also evident is the excess of intermediate-mass objects in the 0.870 pc calculation, with $24\%$ of the stellar population having a mass $\geq 1.0 ~\textup{M}_{\odot}$, compared with $7.3\%$ in the 0.404 pc calculation and $10.4\%$ in the \citeauthor{2005ASSL..327...41C} IMF. This qualitative observation is confirmed by running Kolmogorov-Smirnov tests between the pairs of distributions, which give probabilities of $1 \times 10^{-6}$ and $9 \times 10^{-5}$ of the 0.188 pc and 0.870 pc populations, respectively, being drawn from the same underlying distribution as the 0.404 pc population.  The probability that the 0.188 pc and 0.870 pc populations are drawn from the same underlying distribution is only $2 \times 10^{-9}$. 

We also note that the results of our 0.404 pc calculation and those of the previous 0.404 pc calculation by \citet{2012MNRAS.419.3115B} are statistically indistinguishable, with a $0.149$ probability of being drawn from the same underlying distribution according to a Kolmogorov-Smirnov test. This shows that the mass functions are insensitive to the particular random realisation of the initial velocity field, as we use different seeds to those used in the previous calculation.

\begin{figure*}
	\includegraphics{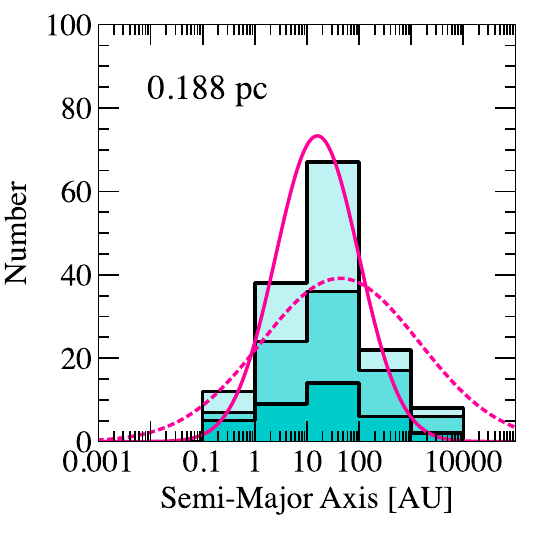}
	\includegraphics{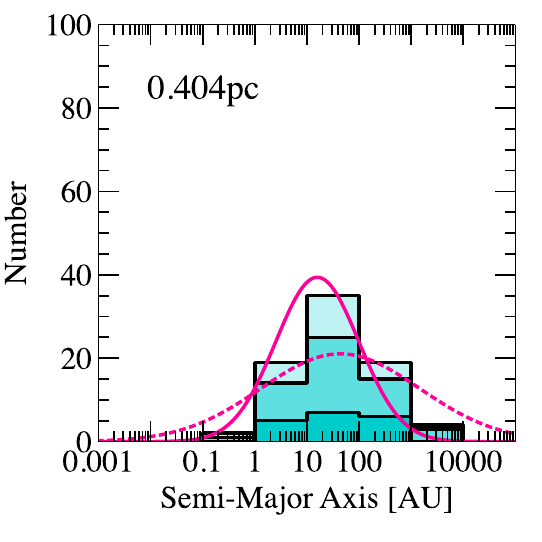}
	\includegraphics{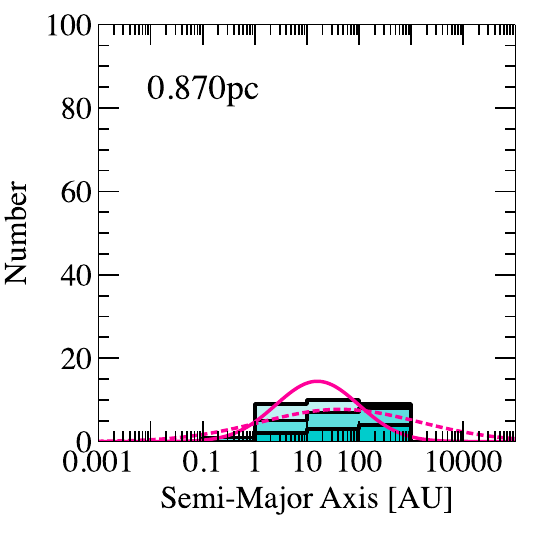}
	\caption{The distribution of semi-major axes in multiple systems with stellar primaries ($M_{1} > 0.1~\textup{M}_{\odot}$) in the each of the calculations at time $t = 1.2~t_{\textup{ff}}$. The light shaded areas show the semi-major axes of binaries; the medium shaded areas show the semi-major axes from triples (two from each triple); and the dark shaded areas show the semi-major axes from quadruple systems. We also plot M-dwarf separation distribution from the M-dwarf survey of \citet{2012ApJ...754...44J} (solid line), as well as the solar-type separation distribution from \citet{2010ApJS..190....1R} (dotted line) in magenta for comparison. There is no significant difference between the distributions, and all are in reasonable agreement with the parameterisations shown. As noted in \citet{2014MNRAS.442..285B}, the results are expected to match the Janson et al. distribution better, as the the simulated systems are primarily low-mass.}
	\label{SMA_hist}
\end{figure*}

It should be noted here that the `IMFs' plotted here are actually protostellar mass functions (PMFs; see \citealt{2010ApJ...716..167M} \citealt{2011ApJ...736...53O}).  As shown by the dark-shaded areas in Fig. \ref{IMF_hist}, a number of objects are still accreting (defined as having an accretion rate $\ge 10^{-7}~\textup{M}_{\odot}~{\rm yr}^{-1}$) by the end of the calculations. \citet{2012MNRAS.419.3115B} found that the distributions of stellar properties evolved such that data taken at any instant of time was representative of the same underlying distribution. Stopping the calculations at $1.2~t_{\textup{ff}}$ therefore does not affect that statistical analysis presented here, as the forms of the distributions are expected to remain constant throughout the duration of the simulation. Therefore, for the remainder of this paper, we refer to the distribution of masses in each cluster as `IMFs' and compare them to observed IMFs, as current methods are unable to accurately determine the PMFs of clusters observationally. 

\begin{figure}
	\includegraphics{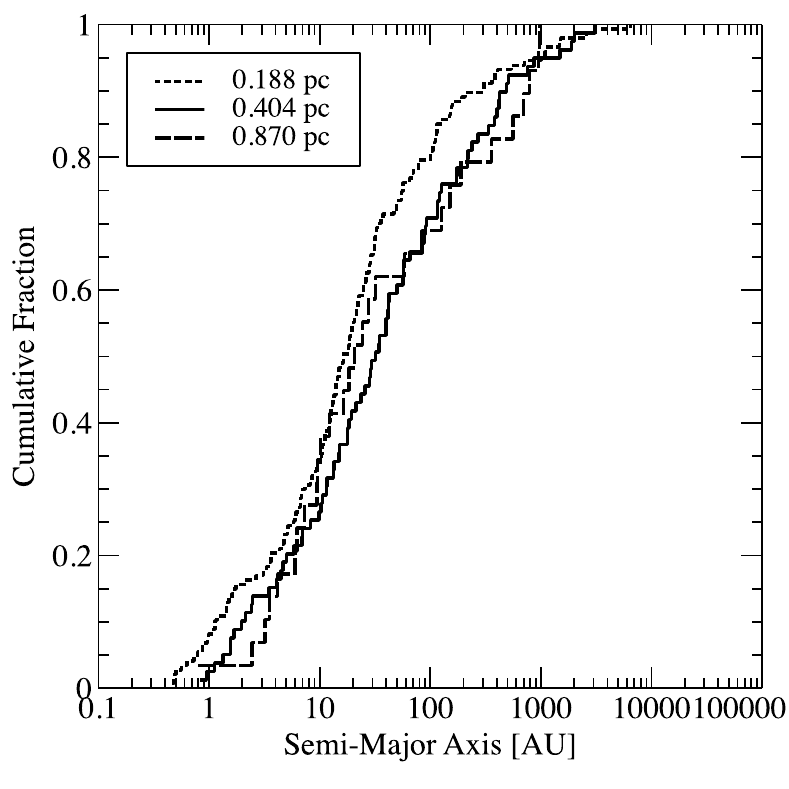}
	\caption{The cumulative distribution of semi-major axes for all multiple systems in the each of the calculations at time $t = 1.2~t_{\textup{ff}}$. Results from the 0.188 pc calculation are shown by the dotted line, the 0.404 pc results by the solid line, and the 0.870 pc results by the dashed line. We include all orbits from binary, triple and quadruple systems. Kolmogrov-Smirnov tests performed on the distributions show that they are statistically indistinguishable.}
	\label{SMA_cum}
\end{figure}

\subsection{Multiplicity as a function of primary mass} \label{results_mult}

To quantify the relative abundance of multiple systems in each calculation, we use the multiplicity fraction as in \citet{2009MNRAS.392.1363B} and \citet{2012MNRAS.419.3115B}, defined as a function of stellar mass by
\begin{equation}
	mf = \frac{B + T + Q}{S + B + T + Q}
\end{equation}
where $S$, $B$, $T$ and $Q$ are the numbers of single stars, binaries, triples and quadruples with primary masses in the same mass range. As discussed in \citet{2012MNRAS.419.3115B}, this method is relatively insensitive to observational incompleteness and subsequent dynamical evolution. 

\begin{figure*}
	\includegraphics{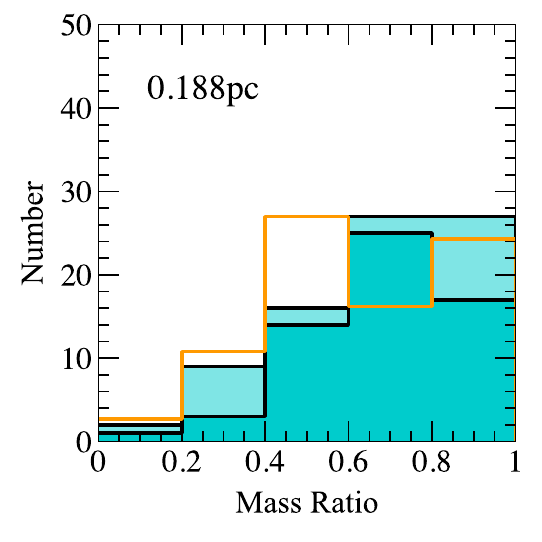}
	\includegraphics{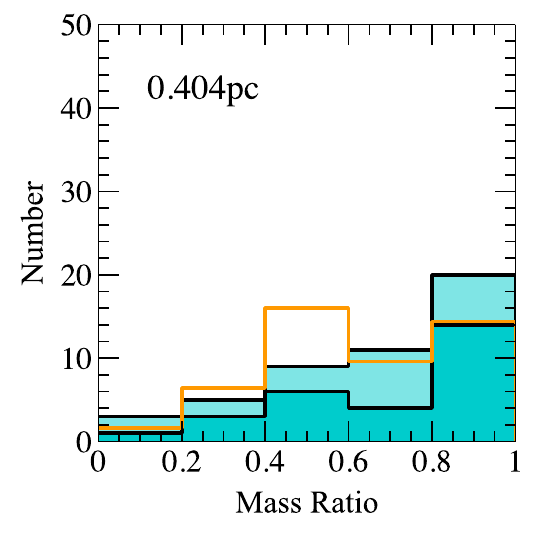}
	\includegraphics{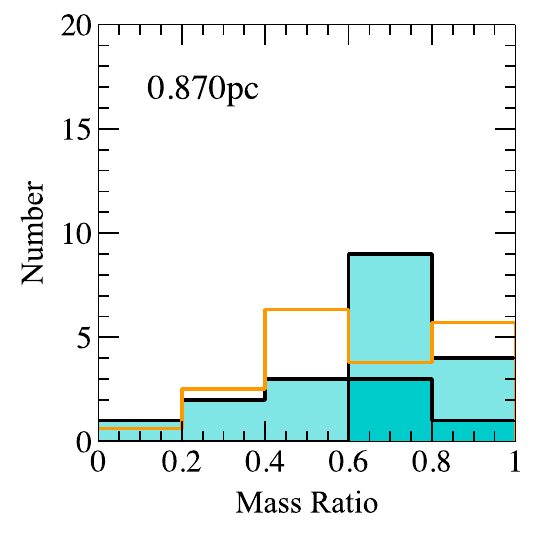}
	\caption{The distribution of mass ratios of stellar pairs in the each of the calculations at time $t = 1.2~t_{\textup{ff}}$. The light shaded areas show the ratios of systems with stellar primaries in the mass range $M_{1} > 0.5~M_{\odot}$, and the dark shaded areas show the ratios of systems with stellar primaries in the mass range $M_{1} = 0.1-0.5~M_{\odot}$. We include the ratios of binaries and binary components of higher-order systems. We also plot the results of \citet{2010ApJS..190....1R} in orange for comparison. The 0.870 pc calculation produces an excess of systems in the 0.6-0.7 range, but there are otherwise no clear differences between the distributions, which are in good agreement with the observations.}
	\label{MR_hist}
\end{figure*}

We use the same algorithm to detect multiple systems as in \citet{2009MNRAS.392.1363B} and \citet{2012MNRAS.419.3115B}, details of which can be found in the former paper. Ambiguities arise when classifying the components of multiples systems, as some binary systems may be part of a triple or quadruple system, and similarly some triple systems may be part of a quadruple system. In this paper, unless otherwise stated, we do not count multiples that are components of higher order systems separately (e.g. a binary system with a third wide companion would only contribute to the number of triples, and not to the number of binaries). We also choose to ignore systems of higher order than quadruples, as such systems are usually dynamically unstable and are therefore unlikely to remain if the calculations were evolved for longer.

\begin{figure}
	\includegraphics{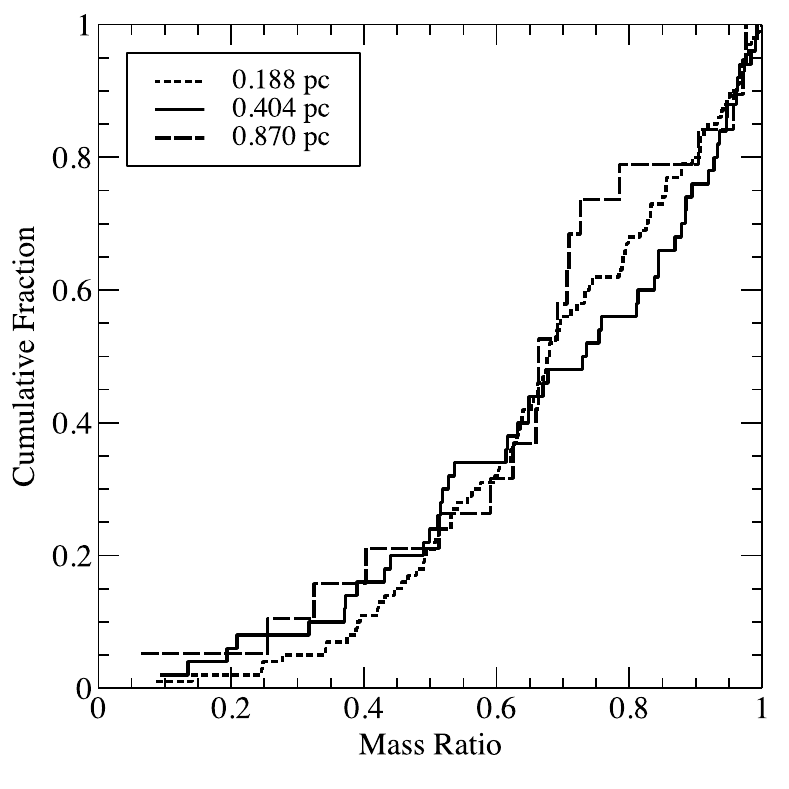}
	\caption{The cumulative distribution of mass ratios in multiple systems in the each of the calculations at time $t = 1.2~t_{\textup{ff}}$. Results from the 0.188 pc calculation are shown by the dotted line, the 0.404 pc results by the solid line, and the 0.870 pc results by the dashed line. We include the ratios of binaries and binary components of higher-order systems. An excess of systems in the 0.7-0.9 range in the 0.870 pc calculation is evident, but the distributions are otherwise in good agreement with each other. Kolmogorov-Smirnov tests performed on the distributions show there are no statistically significant differences between the distributions.}
	\label{MR_cum}
\end{figure}

\citet{2012MNRAS.419.3115B} provided a detailed discussion of the formation and evolution of multiple systems in radiation hydro-dynamical calculations. In the interest of brevity we do not repeat it here, and instead focus on the effects of initial cloud density on the final states of the simulated clusters. The total multiplicities for each of the calculations are 0.31, 0.34 and 0.33 for the 0.188 pc, 0.404 pc and 0.870 pc clouds respectively, each with $1 \sigma$ uncertainties of $\pm 5 \%$. We therefore conclude that there is no significant variation in the overall multiplicity with initial cloud density. 

Fig. \ref{multfrac_all} shows the multiplicity fraction as a function of primary mass for each of the calculations, and compares them with values obtained from various observational surveys (see caption for details). The solid line gives the continuous multiplicity fraction calculated using a lognormal moving average, with the associated $1 \sigma$ and $2 \sigma$ confidence limits shown by the dotted lines, and observations are shown by the squares with error bars. The multiplicity in each calculation increases steadily with primary mass, in agreement with observations of field stars by \citet{2009AJ....137.3358M} (massive stars), \citet{2002AJ....123.1570P} (intermediate-mass stars), \citet{2010ApJS..190....1R} (solar-type stars), \citet{1992ApJ...396..178F} (M-dwarfs) and \citet{2006AJ....132..663B} (very-low-mass (VLM) stars and brown dwarfs).
 
However, whilst the multiplicities of the simulated clusters agree with the values observed for lower stellar masses, all of the calculations exhibit multiplicities higher than observed for solar-type stars. Observations have shown that the multiplicity of young stars is often higher than the field value (\citealt{1993AJ....106.2005G}, \citealt{1995ApJ...443..625S}, \citealt{1999A&amp;A...341..547D}, \citealt{2004A&amp;A...427..651D}, \citealt{2007A&amp;A...476..229D}). Given the young age of the stars formed in the calculations, this provides a plausible explanation for the high values for solar-type stars. It is likely that, as the stellar population ages, some systems will decay through dynamic interactions, reducing the multiplicity.

\subsection{Separation distributions of multiples} \label{results_sep}
Fig. \ref{SMA_hist} shows the distributions of semi-major axes of all multiples systems with stellar primaries (i.e. $M \geq 0.1~\textup{M}_{\odot}$) in each of the three calculations.  Due to the low numbers of VLM systems produced by each of the calculations (19, 2 and 0 in the 0.188 pc, 0.404 pc and 0.870 pc calculations, respectively), we do not discuss their statistical properties.

We compare with results from surveys of M-dwarfs by \citet{2012ApJ...754...44J}, shown by the solid line, and solar-type stars by \citet{2010ApJS..190....1R}, shown by the dotted line. It should be noted here that, although binaries are not assumed to merge until they reach a separation of 0.015 AU, the accretion radii of sink particles in the calculations is significantly larger (0.5 AU), preventing us from accurately modelling any interactions between the stars and gas within this radius, which may be important in the formation of closely-bound systems \citep{2002MNRAS.332L..65B}. There is no obvious difference between each of the distributions, with each calculation's median separation in the range $10-100~\textup{AU}$, similar to observed stellar systems.

Fig. \ref{SMA_cum} displays the cumulative separation distributions for all systems (i.e. both stellar and VLM systems). We note a slight deficit of systems with separations $> 100~\textup{AU}$ in the 0.188 pc calculation, which comprise $\sim 20 \%$ of the total number of multiple systems, compared with $\sim 30 \%$ of the total number of systems in the 0.404 pc and 0.870 pc calculations. This may be due to the dynamical disruption of systems with wide separations, as observed in star-forming regions in Orion (\citealt{1998ApJ...500..825P}; \citealt{1999MNRAS.306..253S}). Kolmogorov-Smirnov tests performed on the distributions find that the 0.188 pc distribution is different from the 0.404 pc calculation at the $2 \sigma$ level. Tests between the other calculations' distributions find them to be statistically indistinguishable. 

The distributions of binary, triple and quadruple systems in all of the calculations are similar, although the numbers of systems are small. There is a noticeable lack of close-binaries, with separations less than 0.5 AU, in each calculation, however this is likely due to the lack of dissipative effects caused by the size of the sink particle accretion radii.

\subsection{Mass-ratio distributions of multiples} \label{results_MR}
Fig. \ref{MR_hist} shows the distribution of mass-ratios of systems with stellar primaries (i.e. $M \geq 0.1~\textup{M}_{\odot}$) in each calculation. We consider only binary systems here, but include binary components of higher-order multiples.  We also plot mass ratio distribution of solar-type binaries from \citet{2010ApJS..190....1R} for comparison.

Once again, the distributions are similar, with no clear peak value except in the case of the 0.188 pc, which produced a large number of systems with mass ratios in the 0.6-0.7 range compared to the other calculations. There is also an absence of systems with ratios less than 0.1 across all of the distributions. This is consistent with the expectations of a `brown dwarf desert', as also noted by \citet{2010ApJS..190....1R}.

Fig. \ref{MR_cum} shows the cumulative distribution of binary mass ratios in each calculation (including VLM binaries). Despite the excess in the number of systems in the 0.6-0.7 range in the 0.188 pc calculation, Kolmogorov-Smirnov tests show that all three of the distributions are statistically indistinguishable from each other.

Overall, the distributions are similar to those presented in \citet{2012MNRAS.419.3115B,2014MNRAS.442..285B}, and are in good agreement with the \citet{2012ApJ...754...44J} and \citet{2010ApJS..190....1R} results. As with the previous section, we do not discuss the properties of VLM systems, as too few were produced.

\section{Discussion} \label{disc}
With the three calculations presented in this paper, we have investigated the effects of changing the initial density of a star-forming molecular cloud on the properties of the resultant stellar populations. Having simulated clouds varying in initial density by two orders of magnitude, we have found evidence of a dependence of the chacteristic stellar mass on the initial cloud density of approximately as $ M_{\rm c} \propto \rho^{-1/5}$, but no significant change in the multiple system properties. We now compare these results to previous theoretical studies and observational surveys. 

\subsection{The dependence of the IMF on density} \label{disc_IMF}
Several studies have put forward theories to explain the apparent universality of the IMF. Each of these theories differs from the others in its predictions about the dependencies of the characteristic stellar mass which, along with the mass spread, defines the IMF. In particular, they all predict a different scaling of the characteristic mass with molecular cloud density. With the results presented in this paper, we are now in a position to compare these predictions. 

\subsubsection{Jeans mass} \label{disc_IMF_Jeans}
When attempting to predict the characteristic (median) mass of the IMF, the thermal Jeans mass is a natural starting point (\citealt{1998MNRAS.301..569L}, \citealt{2005MNRAS.359..211L}). The Jeans length and associated Jeans mass for a gas cloud of uniform temperature and density can be written as
\begin{equation}
	\lambda_{J} = \frac{c_s}{\sqrt{G \rho}} ,
	\label{Jeanslength}
\end{equation}
\begin{equation}
	M_{J} = \frac{4 \pi}{3} \lambda_{J}^3 \rho ,
	\label{Jeansmass_length}
\end{equation}
where $c_s$ is the gas sound speed, and $\rho$ is the gas density. Combining equations \ref{Jeanslength} and \ref{Jeansmass_length} gives the following form for the Jeans mass
\begin{equation}
	M_{J} \propto \frac{c_s^3}{\sqrt{\rho}}.
	\label{Jeansmass}
\end{equation}
For a molecular cloud of total mass $M$, the gas will be divided up into $\sim M/M_{\rm J}$ objects. In the absence of other effects, these gravitationally bound structures will tend to collapse to form protostars with masses equal to $M_{\rm J}$. However, density perturbations, competitive accretion and dynamical interactions between protostars in the same cluster ensure that some stars are able to accrete more material than others, broadening the distribution of masses to create an IMF \citep{1997MNRAS.285..201B}. From this simple analysis it seems clear that the median mass of the IMF should scale with cloud density according to the Jeans mass, which, from equation \ref{Jeansmass}, depends on density as $M_{\rm J} \propto \rho^{-1/2}$. 

In their simplest form, semi-analytic theories of the IMF based on the turbulence-generated structure of molecular clouds also predict a linear scaling of the characteristic stellar mass on the Jeans mass \cite{2002ApJ...576..870P,2008ApJ...684..395H,2009ApJ...702.1428H,2012MNRAS.423.2037H}.  In addition, they predict that the characteristic stellar mass should decrease with increasing levels of turbulence.

Numerical calculations of star formation in molecular clouds confirm a linear dependence of the characteristic stellar mass on the global Jeans mass when the calculations use a barotropic gas equation of state. \citet{2005MNRAS.356.1201B} compared two hydrodynamical simulations of star formation which used identical initial conditions, but which varied in initial density by a factor of 9 (factor of 3 in $M_{\rm J}$), and found that the median mass of the cluster populations varied by a factor of 3 between the calculations. Subsequent calculations by \citet{2006MNRAS.368.1296B} confirmed this scaling, and found that the `knee' of the IMF, i.e. the mass at which the form of the IMF transitions from a flat slope to a steeper slope of the type described by \citet{1955ApJ...121..161S}, was located at the approximate mass scale of the initial Jeans mass of the cloud. 

Calculations performed by \citet{2005A&amp;A...435..611J} using a piecewise polytropic equation of state found that the median mass was dependent on the critical number density, $n_{\rm c}$, at which the dominant form of cooling was assumed to switch from molecular line cooling to thermal cooling, as $M_{\rm c} \propto n_{\rm c}^{-1/2}$. This led them to suggest that the balance between heating and cooling processes, which depend on fundamental physics and chemical abundances and can therefore be derived from universal quantities and constants, determined the characteristic stellar mass, a view supported by \citet{2006MNRAS.368.1296B}. 

\citet{2008ApJ...681..365E} expanded upon this idea, considering star-formation in three distinct environments: dense star-forming cores, the ambient interstellar medium (ISM), and ionising clusters. In dense cores, which most closely resemble the calculations presented in this paper, Elmegreen et al. proposed that molecular cooling resulted in a Jeans mass with a dependence on number density of $M_{\rm J} \propto n^{1/4}$ at the grain-gas coupling point. This relative insensitivity of the Jeans mass to environmental conditions would then give rise to an approximately constant median stellar mass.

Unlike the semi-analytic theories of the IMF based on turbulence-generated structure, the stellar mass functions produced in numerical simulations do not show a strong dependence of the characteristic stellar mass on the level of turbulent driving \cite{2016MNRAS.462.4171B}.  Nor do they provide evidence that the stellar IMF is directly related to the density structure in molecular clouds \citep{2009MNRAS.397..232B,2017MNRAS.465..105L}

\subsubsection{The role of radiative feedback} \label{disc_IMF_Bate}
The first paper to propose radiative heating from protostars as a method of generating a universal IMF was \citet{2009MNRAS.392.1363B}. Similar to a theory put forward by \citet{2006MNRAS.372..143N} to explain the apparent top-heavy IMF of stars observed near Sgr A* in the central region of the galaxy, Bate suggested that radiative feedback from newly-formed, accreting protostars would heat surrounding gas, suppressing fragmentation and increasing the reservoir of gas available for accretion. This would explain the increased characteristic mass observed in the radiative calculation when compared with previous baratropic calculations, and provide a mechanism for the star-formation process to regulate itself.

Results from radiative hydrodynamical calculations of star formation in $50 \textup{M}_{\odot}$ clusters by \citet{2009MNRAS.392.1363B} showed that the inclusion of radiative transfer weakened the dependency of the median stellar mass on the initial cloud density sufficiently to make any variation undetectable with the small number statistics provided by the numerical calculations. In order to explain this behaviour, Bate proposed a simplified analytical model based on the accretion of material of a radiating protostar embedded within a spherically symmetric gas cloud. 

To test the predictions made by \citet{2009MNRAS.392.1363B}, it is beneficial to review the arguments made in that paper. We provide a brief summary of the main points, but encourage interested readers to refer to \citet{2009MNRAS.392.1363B} for a more complete treatment. 

\citet{2009MNRAS.392.1363B} begins by considering the case of gas that is optically thin to infrared radiation and well-coupled to the dust. The gas opacity is assumed to be indepedent of frequency for simplicity. As mentioned in \citet{2009MNRAS.392.1363B}, the result including a wavelength dependent opacity is simple enough to derive, but complicates matters and does not change the conclusions significantly.

Making the above assumptions, the gas temperature, $T$, at a distance, $r$, from a spherically symmetric protostar of luminosity, $L_*$ is obtained from
\begin{equation}
	L_* = 4 \pi r^2 \sigma_{\textup{SB}} T^4,
	\label{luminosity_thin}
\end{equation}
where $\sigma_{\textup{SB}}$ is the Stefan-Boltzmann constant. 

It is possible to express the Jeans length $\lambda_J$ as the distance at which the sound speed of the gas is equal to the escape velocity of the mass enclosed by a sphere of the same radius. This is written as
\begin{equation}
	c^2_{\rm s} = \frac{\mathcal{R}}{\mu} T = \frac{GM}{\lambda_{\rm J}} ,
	\label{Jeanslength_sound}
\end{equation}
where $c_{\rm s}$ is the gas sound speed, $\mathcal{R}$ is the gas constant, and $\mu$ is the mean molecular weight. 

By combining equations \ref{Jeansmass_length}, \ref{luminosity_thin} and \ref{Jeanslength_sound}, we can obtain expressions for the `effective' Jeans mass, $M_{\textup{eff}}$, and Jeans length, $\lambda_{\textup{eff}}$ for a sphere of gas internally heated by a protostar embedded at its centre:
\begin{equation}	
	M_{\textup{eff}} = \rho^{-1/5} L_*^{3/10} \frac{4 \pi}{3}  \left(\frac{3 \mathcal{R}}{4 \pi \mu G} \right)^{6/5} (4 \pi \sigma_{\textup{SB}})^{-3/10},
	\label{Jeansmass_eff}
\end{equation}
\begin{equation}
	\lambda_{\textup{eff}} = \rho^{-2/5} L_*^{1/10} \frac{4 \pi}{3}  \left(\frac{3 \mathcal{R}}{4 \pi \mu G} \right)^{2/5} (4 \pi \sigma_{\textup{SB}})^{-1/10}.
	\label{Jeanslength_eff}
\end{equation}
From these two equations, we see that the effect of the heating from the protostellar luminosity is to modify the thermal Jeans mass, creating an `effective Jeans mass' which depends only weakly on the gas density, with the form $M_{\textup{eff}} \propto \rho^{-1/5}$. There is also a dependence on the protostellar luminosity, of the form $M_{\textup{eff}} \propto L_*^{3/10}$. This will depend on both the properties of the accretion flow, and the structure of the protostar itself.

\citet{2009MNRAS.392.1363B} considered the luminosity of a $0.1 ~\textup{M}_{\odot}$ protostar with a radius of $2 \textup{R}_{\odot}$, taking an accretion rate of $1 \times 10^{-5} ~\textup{M}_{\odot} ~\textup{yr}^{-1}$ from the typical sink particle accretion rates in the calculations, and obtained a protstellar luminosity of $150~\textup{L}_{\odot}$. Assuming an initial mean density of $1.2 \times 10^{-19}~\textup{g}~\textup{cm}^{-3}$ and substituting into equation \ref{Jeansmass_eff}, Bate derived a characteristic mass of $\approx 0.5 ~\textup{M}_{\odot}$.

However, as with the calculations in this paper, the simulations of small stellar clusters performed by \citet{2009MNRAS.392.1363B} did not include radiative feedback from inside the sink accretion radius. Using an accretion radius of $R_\text{acc}=0.5$ AU, and assuming an approximate protostellar radius of $R_*= 2 \textup{R}_{\odot}$, the protostellar luminosity is underestimated by a factor of $R_\text{acc}/R_* \sim 50$. Assuming equation \ref{Jeansmass_eff} holds, this implies that the effective Jeans mass maybe underestimated by up to a factor of $\approx 3$.

This order-of-magnitude calculation assumes that the protostellar luminosity is independent of the mean cloud density. In reality, the luminosity depends upon the mass accretion rate, which will depend on the density of surrounding material, and the protostellar mass-radius relation, which will depend on the entropy content of the protostar. Therefore, to obtain a more accurate expression for the effective Jeans mass, an expression for the protostellar luminosity in terms of the mass accretion rate and protostellar mass-radius ratio is required.

The dominant source of luminosity for newly formed protostars is the conversion of gravitational potential energy via the accretion of mass. The luminosity generated by this process is given by
\begin{equation}
	L_* = \epsilon_{\rm L} \frac{GM_*}{R_*} \dot{M},
	\label{accluminosity}
\end{equation}
where $\epsilon_{\rm L}$ is the efficiency of energy conversion, $M_*$ and $R_*$ are the protostellar mass and radius, and $\dot{M}$ is the mass accretion rate. If we assume that the time-scale for the accretion of the surrounding gas cloud is $\approx \lambda_{\textup{eff}} / c_{\rm s}$, where $c_{\rm s}$ is the sound speed at radius $\lambda_{\textup{eff}}$ from the central protostar, we obtain an approximate accretion rate of
\begin{equation}
	\dot{M} \sim \frac{M_{\textup{eff}} c_{\rm s}}{\lambda_{\textup{eff}}} = \frac{4 \pi \rho \lambda_{\textup{eff}}^2}{3} \left( \frac{\mathcal{R}}{\mu} T \right)^{1/2}.
	\label{accrate}
\end{equation}
Following the same steps as before, this gives an effective Jeans mass of
\begin{equation}
	M_{\textup{eff}} \propto \left( \frac{M_*}{R_*} \right)^{3/7} \rho^{-1/14}.
	\label{Jeansmass_eff_adv}
\end{equation}
The dependence on density is now much weaker, and there is a dependence on the ratio between protostellar mass and radius, which comes from the conversion of gravitational potential energy to luminosity as the protostar accretes. The effect of this second term is difficult to predict, as it will likely depend on the accretion rate as well as the entropy content of the protostar. 

One final point of consideration is the effect of higher optical depths. \citet{2009MNRAS.392.1363B} briefly considered the case where the gas is optically thick to infrared radiation, which results in the transport of radiation becoming diffusive, finding an effective Jeans mass of
\begin{equation}
	M_{\textup{eff}} \propto \rho^{-1/3} L_*^{1/3}.
	\label{Jeansmass_eff_thick}
\end{equation}
The density dependence here is stronger than in the optically thin case, as is the luminosity dependence. \citet{2009MNRAS.392.1363B} did not discuss this case any further, suggesting that the conditions under which it applies (mean number densities $\ge 10^8~\textup{cm}^{-3}$; temperatures $\ge 30~\textup{K}$) are very different to those in local star forming regions.

\subsubsection{The potential impact of protostellar evolution} \label{disc_IMF_Krumholz}
\citet{2011ApJ...743..110K} attempted to extend the analysis of \citet{2009MNRAS.392.1363B} by including the effects of protostellar evolution in order to more accurately determine the accretion luminosity. Similar to \citet{2009MNRAS.392.1363B}, Krumholz proposed that deviations from isothermality are the mechanism by which the characteristic mass scale of the IMF is defined. By relating accretion luminosity to the structure of protostars, Krumholz aimed to link the mechanism by which non-isothermal regions are created to fundamental physics, which would then provide an explanation for the universality of the IMF. 

Again, we review the main arguments of \citet{2011ApJ...743..110K} for the purpose of comparison with the previous study by \citet{2009MNRAS.392.1363B} and with the results of the calculations presented here. 

Similar to \citet{2009MNRAS.392.1363B}, Krumholz begins by considering the minimum mass able to gravitationally collapse to form a new protostar. Krumholz chooses the Bonnor-Ebert mass, which is approximately equal to the Jeans mass, and defined by
\begin{equation}
	M_{\textup{BE}} = 1.18 \sqrt{\left( \frac{k_{\textup{B}} T_{\rm e}}{\mu_{\textup{H}_2} m_{\textup{H}} G} \right)^3 \frac{1}{\rho_{\rm e}}} ,
	\label{BEmass}
\end{equation}
where $T_{\rm e}$ and $\rho_{\rm e}$ are the gas temperature and density at the edge of the sphere containing $M_{\textup{BE}}$. A fraction of this mass will be lost via outflows and jets, such that the final stellar mass will be $M_* = \epsilon_{\rm M} M_{\textup{BE}}$, where $\epsilon_{\rm M}$ is the fraction of collapsed mass accreted by the protostar

To estimate the value of the characteristic mass, the temperature profile of the sphere must be calculated. So long as protostellar heating remains the dominant form of heating, the temperature of the sphere will depend upon the luminosity of the protostar.

As before, proceeding beyond this point requires an estimate of the accretion rate and mass to radius ratio of the protostar. Krumholz assumes an accretion rate $\dot{M} \approx \epsilon_{\rm M} M_{\textup{BE}}/t_{\textup{dyn}}$, based on the dynamic time of the sphere $t_{\textup{dyn}} \approx 1/\sqrt{G \bar{\rho}} = \sqrt{(3-k_{\rho}/(3 G \rho_{\rm e})}$ which, when combined with equation \ref{accluminosity}, results in a protostellar luminosity of 
\begin{equation}
	L_* = \epsilon_{\rm L} \epsilon_{\rm M} \sqrt{\frac{3 G \rho_{\rm e}}{3-k_{\rho}}} M_{\textup{BE}} \psi ,
	\label{luminosity_krumholz}
\end{equation}
where $k_{\rho}$ is defined by the sphere's density profile as $\rho = \rho_{\rm e}(r/R)^{-k_{\rho}}$, and $\psi$ is defined as $\psi \equiv G M_* / R_*$. 

In order to derive the protostellar mass to radius relation, Krumholz considers the collapse of a prestellar core to form an $n = 3/2$ polytrope, supported by deuterium burning. Solving the equations of stellar structure, the following relationship between the protstellar mass and radius can be written
\begin{equation}
	\psi = \left( \frac{T_{\rm n}}{4 \Theta_{\rm c}^3} \right) \frac{E_{\rm G}}{\mu_{\rm i} m_\textup{H}} ,
	\label{MRratio}
\end{equation}
where $T_{\rm n}$ is a normalisation constant, $\mu_{\rm i}$ is the mean molecular weight of ionised gas, $E_{\rm G} = 0.66 \textup{MeV}$ is the Gamow energy and $\Theta_{\rm c}$ is related to the core temperature by $T_{\rm c} \approx E_G / (4 k_{\textup{B}} \Theta_c^3)$. 

By combining equations \ref{BEmass}, \ref{luminosity_krumholz} and \ref{MRratio}, and assuming that $\Theta_c$ is independent of protostellar mass, Krumholz obtains the following expression for the characteristic mass
\begin{equation}
	M_* = m_{\textup{H}} \left( \frac{1.18^{64} 2^{69} 5^{21}}{3^{17} \pi^7} \right) \left( \frac{T_{\rm n}^4 \epsilon_{\rm L}^4 \epsilon_{\rm M}^{13}}{\mu_{\textup{H}_2}^{16} \mu_{\rm i}^4} \right)^{1/9} \left( \frac{\alpha^{16}}{\alpha_{\rm G}^{25}} \right)^{1/18} \Theta_{\rm c}^{-4/3} \left( \frac{P}{P_{\rm P}} \right)^{-1/18}
	\label{stellarmass_krumholz}
\end{equation}
where $\alpha_{\rm G} = G m_{\textup{H}}^2 / (\hbar c )= 5.91 \times 10^{-39}$ is the gravitational fine structure constant for two protons and $P_{\rm P} = c^7 /( \hbar G^2) = 4.63 \times 10^{114}~\textup{dyn}~\textup{cm}^{-2}$ is the Planck pressure. 

As \citet{2011ApJ...743..110K} notes, the predicted characteristic mass depends largely on fundamental constants and dimensionless variables that describe either the geometry of the accretion flow, structure of the protostar or properties of the gas, and are order unity. The only explicit dependence on the system's initial conditions is a very weak dependence on the interstellar pressure. Krumholz cites this as the reason behind the apparently universal form of the observed IMF, as variations in the interstellar pressure over a range of reasonable values produce results that would be indistinguishable observationally.

However, Krumholz relies upon the assumption that, following the second-stage collapse of a hydrostatic core, the resultant protostar is supported by thermal pressure generated by deuterium burning, in order to link the internal structure of the protostar to fundamental physical constants and processes. In fact, deuterium burning is not expected to begin until later in the protostar's lifetime (\citealt{2000ApJ...542L.119C}), by which time most star-formation in the vicinity will have ceased. Instead, protostars are supported by thermal pressure generated via gravitational contraction during the period immediately following their formation, which results in internal structures that are heavily dependant on their accretion histories and thermodynamic content (\citealt{1997ApJ...475..770H,1999MNRAS.310..360T,2002A&amp;A...382..563B}). As such, although this simplified model highlights the importance of protostellar evolution in correctly modelling protostellar accretion luminosity and determining the form of the IMF, its predictions should be treated with caution.

\subsubsection{Comparison between analytic and numerical results} \label{disc_IMF_sum}
The numerical calculations presented in this paper produce stellar mass functions whose characteristic mass scales with density as $M_c \propto \rho^{-1/5}$.  This scaling with density is much weaker than that expected from a simple Jeans mass argument, or from the basic semi-analytic turbulent models of the origin of the IMF. Instead, the numerical scaling is in agreement with the simple optically thin description (equation \ref{Jeansmass_eff}) given by \citet{2009MNRAS.392.1363B}.

However, the calculations performed for this paper use sink particles with accretion radii much larger than typical protostellar radii and they do not emit radiation that would be generated from within the accretion radii.  This leads to less radiative heating than would actually be produced in reality, and means that the radiative heating that is produced may not scale in the same way as if the protostars were modelled accurately. Generally speaking, more radiative heating would be expected to decrease the amount of fragmentation and increase the characteristic stellar mass.  However, if accretion is episodic this may not necessarily be the case.  \citet{2015MNRAS.447.1550L} considered fragmentation of low-mass pre-stellar cores and found that with continuous radiative feedback brown dwarfs were under-produced and multiple star properties were in poor agreement with observations.  However, when they treated the feedback as being episodic they obtained more realistic stellar properties that were similar to those obtained without including radiative feedback at all.

Because of the incomplete protostellar heating, the disagreement between our results and the predictions of the more advanced treatments of \citet{2009MNRAS.392.1363B} (equation \ref{Jeansmass_eff_adv}) and \citet{2011ApJ...743..110K} (equation \ref{stellarmass_krumholz}) is, perhaps, unsurprising. These models make estimates for the luminosities of the protostars that depend on assumptions about the rate of mass accretion during protostar formation, and upon knowledge of the protostellar structure to calculate the accretion luminosity of protostars, both of which are not resolved by our calculations. As such, we are unable to rule out either of these more advanced theories based on the results of this paper alone. We hope to remove this limitation in future work.

\subsection{Comparison with observations} \label{disc_obs}
As with previous calculations of this scale \citep{2012MNRAS.419.3115B,2014MNRAS.442..285B}, the number of stars and brown dwarfs produced by each of the three clusters presented is sufficient for us to make comparisons between the distributions of properties in the simulated clusters and those found in observational surveys. 

It is now widely accepted that the number of stars produced in star-forming regions is higher than the number of brown dwarfs (\citealt{2003PASP..115..763C}; \citealt{2007AJ....133.1321G}; \citealt{2007ApJS..173..104L}; \citealt{2008ApJ...683L.183A}). The results of our calculations are in agreement with this, with brown dwarfs making up 42\%, 31\%, 24\% of the stellar populations in the 0.188 pc, 0.404 pc, and 0.870 pc calculations respectively. We also find that the median mass does vary with initial cloud density, albeit more weakly than would be expected from the simplest Jeans-mass arguments. While many past observations have suggested that the median stellar mass is approximately constant \citep{2010ARA&amp;A..48..339B}, there is some evidence for variation with environmental conditions. 

Low-density star-forming regions, such as the nearby Taurus-Auriga cloud complex, have been observed to contain an apparent excess of solar-type stars when compared to common parameterisations of the IMF \citep{2003ApJ...590..348L}. Of the calculations presented in this paper, the low-density 0.870 pc calculation is the most comparable. As in Taurus-Auriga, we find an excess of intermediate-mass stars (see Fig. \ref{IMF_hist}) in the cluster. \citet{2004A&amp;A...419..543G} suggested the excess of solar-type stars may be due to dynamic interactions between small groups of protostars in star-forming cores early in the formation process. Ejections of brown dwarfs and low-mass stars would deplete the mass of the core by a negligible amount, leaving the remaining objects with a larger proportion of the gas reservoir from which to accrete, increasing their final masses. Star-formation in our low-density 0.870 pc calculation occurs in small isolated groups similar to those described by \citet{2004A&amp;A...419..543G} (see Fig. \ref{denstemp}). These systems do evolve through dynamic ejections in a manner similar to that proposed by \citet{2004A&amp;A...419..543G}.
 
Observations have also found evidence of bottom-heavy mass distributions in the inner regions of early-type galaxies, with median masses that decrease with increased velocity dispersions (\citealt{2010ApJ...709.1195T}; \citealt{2012ApJ...753L..32S}; \citealt{2012ApJ...760...71C}; \citealt{2017ApJ...846..166L}). \citet{2012ApJ...760...71C} suggest several explanations for this, including increased pressure in the interstellar medium, and high star-formation rates.  Direct comparisons with our calculations are difficult because the initial conditions for the stellar populations in the inner regions of early-type galaxies are not known.  However, if these stars were predominantly produced from high-density molecular gas, the fact that we obtain lower characteristic stellar masses in regions of high density may help explain the origin of such a bottom-heavy mass function.

\section{Conclusions} \label{conc}
In this paper, we have presented results from three radiation hydrodynamical calculations of star formation in clusters that resolve masses down to the opacity limit for fragmentation. 

Each of the calculations began with the same basic initial conditions, but the mean density was varied by up to a factor of 100 between the clouds by changing their initial radii. The calculations produced 474, 233 and 87 stars and brown dwarfs, which we modelled using sink particles with accretion radii of 0.5 AU. The large number of objects formed allows us to examine the statistical properties of the clusters and compare them to observational surveys of nearby star-forming regions, and previous analytical theories.

Our conclusions for how stellar properties in star-forming clusters depend on the initial cloud density are as follows.

(i) We find that including the effects of radiative feedback in the calculations produces clusters whose median stellar masses exhibit a weaker dependency on the initial cloud density than the thermal Jeans mass, confirming the results of previous calculations by \citet{2009MNRAS.392.1363B}.

(ii) We observe a dependency of the characteristic (median) stellar mass on density approximately of the form $M_{\rm c} \propto \rho^{-1/5}$, in contrast to similar, but much smaller, calculations performed by \citet{2009MNRAS.392.1363B}, which found no significant variation in the median mass between calculations of different initial densities.

(iii) The dependency of the median mass on initial density that we find is consistent with the basic treatment of the optically thin case proposed by \citet{2009MNRAS.392.1363B} ($M_{\rm c} \propto \rho^{-1/5}$). 

(iv) The $M_{\rm c} \propto \rho^{-1/5}$ scaling is much stronger than that predicted by the extended models of \citet{2009MNRAS.392.1363B} and \citet{2011ApJ...743..110K} that try to take into the luminosity of accreting protostars. However, our numerical calculations use sink particles that exclude the luminosity that would be generated from within the 0.5~AU accretion radii. Future calculations that include this omitted luminosity will be required to definitively test these extended models.

(v) The multiple system properties of the clusters display little variation and are generally in good agreement with the properties observed for Galactic stellar populations.  All of our calculations exhibit multiplicities higher than those observed for solar-type field stars. This is consistent with surveys of young stars in nearby star-forming regions (see \citealt{1993AJ....106.2005G}, \citealt{1995ApJ...443..625S}, \citealt{1999A&amp;A...341..547D}, \citealt{2004A&amp;A...427..651D}, \citealt{2007A&amp;A...476..229D}).  We also find possible evidence for a weak dependence of the number of wide binaries/multiples on density, which is consistent with wide systems being preferentially disrupted, although the effect is not statistically robust. 

(vi) The lowest density calculation (0.870 pc) exhibits comparable structures to those observed in the Taurus-Auriga star-forming region, and produces a similar stellar mass distribution with an excess of solar-type stars. This suggests that the apparently top-heavy IMF observed in Taurus-Auriga may be due to the region's low density.

(vii) If the characteristic stellar mass does decrease with increasing cloud density, this may help to explain the bottom-heavy stellar mass functions that have been inferred from observations of the inner regions of early-type galaxies.

\section*{Acknowledgements}

This work was supported by the European Research Council under the European Commission's Seventh Framework Programme (FP7/2007-2013 Grant Agreement No. 339248).  The calculations discussed in this paper were performed on the DiRAC Complexity system, operated by the University of Leicester IT Services, which forms part of the STFC DiRAC HPC Facility (www.dirac.ac.uk), and the University of Exeter Supercomputer, Isca. The former equipment was funded by BIS National E-Infrastructure capital grant ST/K000373/1 and STFC DiRAC Operations grants ST/K0003259/1 and ST/M006948/1. DiRAC is part of the National E-Infrastructure.





\bibliographystyle{mnras}
\bibliography{clusters} 


\bsp	
\label{lastpage}
\end{document}